\documentclass[12pt]{article}
\usepackage{putex}
\usepackage{graphicx}
\usepackage{epstopdf}
\usepackage{subfigure}

\def\re{\operatorname{Re}}
\def\half{{{1 \over 2}}}
\providecommand{\abs}[1]{\lvert#1\rvert}

\begin{document}

\preprint{hep-th/0612065 \\ PUPT-2218}

\institution{PU}{Joseph Henry Laboratories, Princeton University, Princeton, NJ 08544}

\title{Low-lying gravitational modes in the scalar sector of the global $AdS_{4}$ black hole}

\authors{Georgios Michalogiorgakis and Silviu S. Pufu}

\abstract{We compute the quasinormal frequencies corresponding to the scalar sector of gravitational perturbations in the four-dimensional AdS-Schwarzschild black hole by using the master field formalism of \cite{Kodama:2003jz}.  We argue that the non-deformation of the boundary metric favors a Robin boundary condition on the master field over the usual Dirichlet boundary condition mostly used in the literature.  Using this Robin boundary condition we find a family of low-lying modes, whose frequencies match closely with predictions  from linearized hydrodynamics on the boundary.  In addition to the low-lying modes, we also see the usual sequence of modes with frequencies almost following an arithmetic progression.}

\PACS{}
\date{December 2006}

\maketitle
\tableofcontents

\section{Introduction}\label{INTRO}

Recently, a lot of attention has been devoted to the study of quasinormal perturbations in asymptotically anti-de Sitter (AdS) backgrounds.  The first quasinormal mode (QNM) computation in AdS space was done in \cite{Chan:1996yk} for a conformally invariant scalar field, and then the problem was solved in \cite{Horowitz:1999jd} for any minimally-coupled scalar field in dimensions $d=4$, $5$, and $7$.  The gravitational perturbations of global $AdS_4$-Schwarzschild, which is what we are interested in, have been computed for the first time in \cite{Cardoso:2001bb}.  Since then, various properties and generalizations of these QNM's have been considered, such as asymptotic relations \cite{Natario:2004jd, Musiri:2005ev}, different anti-de Sitter backgrounds \cite{Cardoso:2001vs, Kovtun:2005ev}, or other boundary conditions \cite{Moss:2001ga}.

We will phrase the problem in terms of the master field formalism that was developed by Kodama and Ishibashi in \cite{Kodama:2003jz}.  Based on previous ideas developed by Regge, Wheeler and Zerilli \cite{Regge57,Zerilli:1970se} that were further extended in \cite{Cardoso:2001bb,Cardoso:2001vs,Kodama:2000fa},  Kodama and Ishibashi developed this formalism to decouple the linearized Einstein equations for the gravitational perturbations of the global AdS-Schwarzschild in a gauge-invariant way and in any number of dimensions. In general, the perturbations in $AdS_d$ can be divided into tensor, vector, and scalar perturbations, depending on whether they correspond to expansions in tensor, vector, or scalar spherical harmonics on the $S^{d-2}$ section of $AdS_d$.  The basic idea then is that we can express each of these perturbations in terms of a master field $\Phi$ and the appropriate spherical harmonics, and all we have to do is solve a certain differential equation satisfied by $\Phi$.  This equation will in general depend on both the perturbation type we are considering and on the number of dimensions.

We will restrict our attention to the scalar sector of the perturbations in $d=4$, where most QNM-related computations in the literature use a Dirichlet boundary condition on the master field $\Phi$ near the boundary of AdS.  The purpose of this paper is to comment on the choice of this boundary condition, and to suggest that a Robin boundary condition\footnote{A Robin boundary condition specifies a linear combination of a function and its derivative at the boundary.} would be more appropriate, especially from the point of view of the AdS/CFT duality (\cite{Maldacena:1997re, Gubser:1998bc, Witten:1998qj}; for a review, see \cite{Aharony:1999ti}).  It follows from the AdS/CFT dictionary that a natural expectation is to demand that the perturbations do not deform the metric on the boundary of AdS, and this condition in turn determines the asymptotic behavior of the master field at the boundary.  While having no boundary deformations amounts in other similar situations to imposing a Dirichlet boundary condition on $\Phi$ at the boundary, this is \emph{not} the case for the scalar sector of gravitational perturbations in $AdS_4$, where a Robin boundary condition is required (see section~\ref{BDYCOND}).\footnote{We can anticipate some trouble in the scalar sector of $AdS_4$ perturbations just by looking at the general large $\rho$ dependence of the master field $\Phi$ for any kind of perturbations and in any number of dimensions.  In general, $\Phi$ satisfies a second order differential equation whose linearly independent solutions behave like $\rho^{{d-6 \over 2}+j}$ and $\rho^{{4-d \over 2}-j}$, where $j = 0$, $1$, or $2$ for scalar, vector, or tensor perturbations, respectively.  The scalar perturbations in $d=4$ are the first ones for which the behavior of the first family is subleading to the behavior of the second family.}  Using the Robin boundary condition proposed in section~\ref{BDYCOND}, we find a family of low-lying modes that were not seen when a Dirichlet boundary condition was used instead (see for example \cite{Cardoso:2001bb}).  In addition to the low-lying modes, we also find a tower of modes that is similar to the tower of modes found in \cite{Cardoso:2001bb}.  For details on what makes the low-lying modes different, or for what other differences we find between our QNM's and the ones computed in \cite{Cardoso:2001bb}, see section~\ref{NUMERICS}.  It is important to note that our Robin boundary condition doesn't affect the vector gravitational perturbations in $AdS_4$, because in this case the Dirichlet boundary condition is still appropriate.
  
A good check on the values of the low-lying quasinormal frequencies comes from a linearized hydrodynamics approximation on $S^2 \times \mathbf{R}$.  The rationale of this approach lies in the observation that since M-theory on  $AdS_4$-Schwarzschild$\times S^7$ is dual to a thermal CFT on the boundary, some QNM's should correspond to hydrodynamic modes of the thermal CFT.  This idea has been developed in several interesting papers:  in \cite{Son:2002sd,Policastro:2002se,Policastro:2002tn} it was shown that the quasinormal frequencies should correspond to poles of the correlation functions on the field theory side, and in \cite{Policastro:2002tn, Kovtun:2005ev,Kovtun:2006pf} this result was checked by explicit numerical computations.  We follow the approach in \cite{Friess:2006kw}, where it was noted that the low-lying scalar and vector modes in \emph{five}-dimensional AdS-Schwarzschild can be computed through a linearized hydrodynamics approximation.  Extending the argument given in \cite{Friess:2006kw} to any dimension, we derive an approximate formula for the low-lying scalar and vector modes in $AdS_d$.  We find excellent agreement between the numerically found low-lying modes (using our Robin boundary condition) and the linearized hydrodynamics prediction in $d=4$.

The paper is organized as follows:  in section~\ref{SETUP} we present an overview of the general setup of our calculation,  in section~\ref{BDYCOND} we comment on the choice of boundary conditions and derive the boundary asymptotics for the master field $\Phi$, in section~\ref{NUMERICS} we show the results of our numerical computation of the quasinormal frequencies of the global $AdS_4$-Schwarzschild solution, and finally, in section~\ref{HYDRO} we compare our results to what one would expect from the analysis of linearized hydrodynamics of a conformal plasma on $S^2 \times \mathbf{R}$.

\section{Setup of the calculation}\label{SETUP}

In this section we briefly review the setup of our calculation.  The global $AdS_4$-Schwarzschild black hole solution is given by
 \eqn{Metric}{
  ds^2 = -\left(1-{\rho_0 \over \rho} + {\rho^2 \over L^2} \right) d\tau^2 + {d\rho^2 \over 1-{\rho_0 \over \rho} + {\rho^2 \over L^2}} + \rho^2 d\Omega_2^2\,,
 }
where $d \Omega_2^2$ is the standard metric on the unit $S^2$,
 \eqn{STwoMetric}{
  d \Omega_2^2 = \gamma_{ij} dy^i dy^j = d \theta^2 + \sin^2 \theta d \phi^2\,.
 }
This metric is a solution to the Einstein equations that follow from the action
 \eqn{EinsteinAction}{
  S = {1\over 2 \kappa^2} \int d^4 x \sqrt{g} \left(R + {6\over L^2}\right)\,.
 }
The horizon radius of the black hole solution \eqref{Metric} is then the positive root of the equation
 \eqn{rhoHDef}{
  \rho_0 = \rho_H \left(1 + {\rho_H^2 \over L^2}\right)\,.
 }
For future reference, the mass, entropy, and Hawking temperature of this black hole solution are:
 \eqn{MET}{
  M = {4 \pi \rho_0 \over \kappa^{2} } \qquad S = {8 \pi^2 \rho_H^2 \over \kappa^{2}} \qquad T = {1 + 3 \rho_H^2 /L^2 \over 4 \pi \rho_H} \,.
 }
We are interested in linear perturbations of the background metric \eqref{Metric}, of the form $g_{ab} + \delta g_{ab}$, that satisfy the linearized Einstein equations following from \eno{EinsteinAction}.  The boundary conditions satisfied by these perturbations will be discussed in section~\ref{BDYCOND}.

The linearized equations satisfied by the perturbations can be solved by separation of variables.  We assume $\delta g_{ab} \sim e^{-i \omega \tau} \Phi(\rho) S_{ab} (\theta, \phi)$, where the functions $S_{ab}$ depend only on the angular variables on $S^2$, and can be written in terms of the spherical harmonics $Y_{lm} (\theta, \phi)$ and generalizations thereof.  The exact equations describing the scalar, vector, and tensor perturbations of the $d$-dimensional AdS-Schwarzschild background can be found in \cite{Kodama:2003jz}.  As discussed previously, we will only focus on the scalar perturbations in the case $d=4$.  Using the notation in \cite{Friess:2006kw}, we split the coordinates $y^{a} = (\tau, \rho, \theta, \phi)$ into $y^\alpha = (\tau, \rho)$ and $y^i = (\theta, \phi)$.  We denote by $\nabla_i$ the covariant derivative with respect to the metric \eqref{STwoMetric} on $S^2$, and by $D_\alpha$ the covariant derivatives with respect to the two-dimensional metric
 \eqn{OrbitMetric}{
  ds_2^2 = -f d\tau^2 + {1 \over f} d\rho^2 \qquad f = 1-{\rho_0 \over \rho} + {\rho^2 \over L^2}\,.
 }
The equations describing the scalar perturbations then read:
 \eqn[c]{ScalarPerturb}{
  \delta g_{\alpha\beta} = f_{\alpha\beta} \,
    \mathbb{S}(\theta,\phi) \qquad
   \delta g_{\alpha i} = \rho f_\alpha \,
     \mathbb{S}_i(\theta,\phi)  \cr
   \delta g_{ij} = 2 \rho^2 \left[ H_L(\tau,\rho) \, \gamma_{ij} \,
    \mathbb{S}(\theta,\phi) + H_T(\tau,\rho) \,
    \mathbb{S}_{ij}(\theta,\phi) \right]
 }
 \eqn{ScalarSDefs}{
  \mathbb{S}_i = -{1 \over k_S} \partial_i \mathbb{S} \qquad
   \mathbb{S}_{ij} = {1\over k_S^2} \nabla_i \partial_j \mathbb{S} + {1\over 2} \gamma_{ij} \mathbb{S}
 }
 \eqn{ScalarParams}{
  H = m + 3 w \qquad w = {\rho_0 \over \rho } \qquad m = k_S^2 - 2
 }
 \eqn{ScalarPhiDefs}{
   X_\alpha = {\rho \over k_S}\left(f_\alpha + {\rho \over k_S} \partial_\alpha H_T\right) \cr
   F_{\alpha \beta} = f_{\alpha \beta} + D_\alpha X_\beta + D_\beta X_\alpha \cr
   F = H_L + {1 \over 2} H_T + {1 \over \rho} \left(\partial^\alpha \rho \right) X_\alpha
 }
 \eqn{ScalarPhiProperties}{
  F^{\alpha}_{\phantom{\alpha}\alpha} = 0 \qquad D^\alpha F_{\alpha\beta} = 2 \partial_{\beta} F
 }
 \eqn{ScalarPhi}{
    F_{\alpha \beta} = {1 \over H} \left( D_\alpha \partial_\beta \left(\rho H \Phi\right) - \half g_{\alpha \beta} D_\gamma \partial^\gamma \left(\rho H \Phi\right)\right)
 }
 \eqn{ScalarEOM}{
  \left( D_\alpha \partial^\alpha -
    {V_S(\rho) \over f} \right) \Phi = 0
 }
 \eqn{ScalarPotential}{
  V_S(\rho) &= { f \over \rho^2 H^2} \left[m^3 + m^2 \left( 2 + 3 w\right) + 9m w^2 + 9 w^2 \left(2f + 3w - 2 \right)\right] \,.
 }
where $\mathbb{S}$ denotes any of the spherical harmonics $Y_{lm}$ on $S^2$, and $k_S^2$ is the corresponding eigenvalue of the laplacian:
 \eqn{SphericalDef}{
  \left( \nabla_i \partial^i + k_S^2 \right) \mathbb{S} = 0 \qquad \mathbb{S}(\theta, \phi) = Y_{lm}(\theta, \phi) \qquad k_S^2 = l(l+1)\,.
 }
It is worth noting that the above master field formulation is gauge invariant.  So equations \eqref{ScalarPerturb}--\eqref{ScalarPotential} don't determine the perturbations $\delta g_{ab}$ uniquely:  there is an implicit freedom of choosing four of these functions through a gauge transformation of the form $\delta g_{ab} \to \delta g_{ab} + \nabla_a v_b + \nabla_b v_a$, where this time $\nabla_a$ denotes the covariant derivative with respect to the full four-dimensional metric \eqref{Metric}, and $v_a$ are arbitrary functions.  A small discussion of our gauge choice and the residual gauge freedom is included in the Appendix.

\section{Boundary conditions}
\label{BDYCOND}

\subsection{Boundary conditions at $\rho = \infty$}
\label{BOUNDARY}

The question of what boundary conditions one should impose on the master field $\Phi$ at the boundary of AdS does not have a well-established answer:  most of the previous authors have set $\Phi(\infty) = 0$ (see, for example, \cite{Cardoso:2003cj, Konoplya:2003dd, Natario:2004jd, Musiri:2005ev}), but other boundary conditions have also been used (see, for example, \cite{Moss:2001ga}).\footnote{It is not clear to us how the boundary condition that we find is related to the one proposed in \cite{Moss:2001ga}.}  As we shall see below, the AdS/CFT dictionary relating perturbations and expectation values of operators in the dual field theory might help clarify this point.

From the AdS/CFT perspective, there are two independent behaviors of the metric perturbations $\delta g_{ab}$ at large $\rho$:  $\delta g_{ab} \sim \rho^2$, which corresponds to a deformation of the boundary metric, and $\delta g_{ab} \sim 1/\rho$, which corresponds to a non-zero VEV of the stress-energy tensor in the boundary theory.  In defining the quasinormal frequencies it is sensible to require that the metric perturbations \emph{do not} change the boundary metric, so they only produce a non-zero VEV of the stress-energy tensor $\langle T_{ab} \rangle$ of the thermal plasma on the boundary.  This prescription is equivalent to requiring the quasinormal frequencies to correspond exactly to the poles of the correlation functions in the strongly coupled dual CFT in the planar limit (see for example \cite{Son:2002sd}).  With this in mind, the significant challenge is to find the relation between the asymptotic behaviors of $\delta g_{ab}$ and $\Phi$, which is what we'll now turn to.

At large $\rho$, the master equation \eqref{ScalarEOM} takes the form
 \eqn{FarFieldEq}{
  \left[{\Omega^2 \over L^2} + {\rho^2 \over L^4} \partial_\rho \rho^2 \partial_\rho - {k_S^2 \over L^2} - {18 \rho_0^2 \over L^4} {1\over (k_S^2-2)^2} \right] \Phi_{\rm far} = 0\,,
 }
where we have assumed $e^{-i \omega \tau}$ behavior and denoted $\omega = \Omega/L$.  Being a second order differential equation, equation \eqref{FarFieldEq} has two linearly independent solutions.  Their asymptotic behaviors at large $\rho$ are given by:
 \eqn{PhiFar}{
  \Phi_{\rm far} (\rho) &= e^{-i \Omega \tau/L} \left[\varphi^{(0)}  + {\cal O}\left({L^2\over \rho^2}\right) \right] \quad \text{ and } \cr  \Phi_{\rm far} (\rho) &= e^{-i \Omega \tau/L} \left[\varphi^{(1)} {L \over \rho}  + {\cal O}\left({L^3\over \rho^3}\right) \right]\,.
 }

As noted earlier, the boundary condition that has been mostly used in the literature is $\varphi^{(0)} = 0$.  As we shall see shortly, this condition is not consistent with the idea that $\delta g_{ab} \sim 1/\rho$ is the only behavior allowed.  To argue this, we choose to work in axial gauge ($\delta g_{\rho a} = 0$), and we derive the boundary condition on $\Phi$ required by $\delta g_{ab} \sim 1/\rho$.  While we include a detailed and more complete derivation in the Appendix, we now present the simplest way of arriving at the proposed boundary condition.

Setting $L=1$, we can plug \eqref{PhiFar} into equation \eqref{ScalarPhi} and obtain, for the $F_{\tau\rho}$ component
 \eqn{Ftaurho}{
  F_{\tau\rho} = i e^{-i \Omega \tau} \left(\varphi^{(1)} + {3 \rho_0 \varphi^{(0)}  \over k_S^2 - 2}\right) {1 \over \rho} + {\cal O}\left({1\over \rho^2} \right)\; .
 }  
Using axial gauge and, as discussed above, assuming $\delta g_{ab} \sim 1/\rho$, we have:
 \eqn{asympt}{
  f_{\tau\rho} = 0 \qquad f_{\rho\rho} &= 0 \qquad f_{\rho} = 0\cr
  H_L = {A_L^{(3)} e^{-i \Omega \tau} \over \rho^3} + {\cal O}\left({1\over \rho^4} \right) &\qquad
  H_T = {A_T^{(3)} e^{-i \Omega \tau} \over \rho^3} + {\cal O}\left({1\over \rho^4} \right)\cr
  f_{\tau\tau} = {B^{(1)} e^{-i \Omega \tau} \over \rho} + {\cal O}\left({1\over \rho^2} \right) &\qquad
  f_{\tau} = {C^{(2)} e^{-i \Omega \tau} \over \rho^2} + {\cal O}\left({1\over \rho^3} \right)\,.
 }
By using \eqref{ScalarPhiDefs} we can compute
 \eqn{Ftaurhoagain}{
  F_{\tau\rho} = e^{-i \Omega \tau} {-3 C^{(2)} k_S + 6 i A_T^{(3)} \Omega \over k_S^2} {1 \over \rho^2} + {\cal O}\left({1\over \rho^3} \right)\,.
 }
This means that axial gauge and $\delta g_{ab} \sim 1/\rho$ force $F_{\tau\rho}$ to behave as $1/\rho^2$.  By comparing this behavior to the general expectation \eqref{Ftaurho}, we conclude that the $1/\rho$ term in \eqref{Ftaurho} must vanish:
 \eqn{BdyCond}{
  \varphi^{(1)} + {3 \rho_0 \varphi^{(0)} \over k_S^2 - 2}  = 0\,.
 }
Thus we obtain a Robin boundary condition, involving the master field and its derivative. 
\subsection{Boundary conditions at the horizon}
\label{HORIZON}

In contrast to the large $\rho$ boundary conditions whose derivation was somewhat subtle and tedious, the horizon boundary conditions are straightforward, being based on the requirement that classical horizons don't radiate.  So in appropriate coordinates, the perturbations near the horizon should take the form of an infalling wave.  To make this explicit, we define the standard ``tortoise'' coordinate by
 \eqn{Tortoise}{
  r_* = \int {d \rho \over f(\rho)}\,,
 }
which puts the master equation \eqref{ScalarEOM} into the form
 \eqn{NearHorEqn}{
  \left[-\partial_\tau^2 + \partial_{r_*}^2 - V_S(\rho)\right] \Phi = 0\,.
 }
Here, $\rho \to \rho_H$ corresponds to $r_* \to -\infty$.  Noticing that $V_S (\rho_H) = 0$, we can immediately see that the near horizon behavior of the two linearly independent solutions to the master equation are $e^{-i \Omega (\tau \pm r_*)/L}$:
 \eqn{PhiNear}{
  \Phi_{\rm near} (\rho) = U e^{-i \Omega (\tau + r_*)/L} + V e^{-i \Omega (\tau - r_*)/L}\,.
 }
The infalling boundary condition then means setting $V = 0$.

\section{Numerical solutions}
\label{NUMERICS}

\subsection{Change of variables}
\label{CHANGEOFVARS}

In order to solve the master equation \eqref{ScalarEOM} numerically, it is convenient to recast it in terms of a different field $\psi(y)$, defined by factoring out the near horizon behavior of the master field $\Phi(\rho)$:
 \eqn{psiDef}{
  \Phi = e^{-i \Omega (\tau + r_*)/L} \psi(y) \qquad y = 1- {\rho_H \over \rho}\,.
 }
Setting $L=1$, we can plug this ansatz into the master equation \eqref{ScalarEOM} to obtain the differential equation satisfied by $\psi$.  We obtain
 \eqn{psiEOM}{
  \left[s(y) \partial_y^2 + t(y) \partial_y + u(y)\right] \psi(y) = 0\,,
 }
where
 \eqn{stuDef}{
  s(y) &= K(y) (1-y)^4 f^2\cr
  t(y) &= K(y) \left[(1-y)^2 f {\partial \over \partial y} \left[(1-y)^2 f \right] - 2 i \Omega \rho_H (1-y)^2 f \right]\cr
  u(y) &= -K(y) \rho_H^2 V_S\cr
  K(y) &= {1\over y} \left[1 + k_S^2 + 3 \rho_H^2 - 3 y (1+ \rho_H^2) \right]^2\,.
 }
Here, $K(y)$ has been chosen so that $s(y)$, $t(y)$, and $u(y)$ are polynomial expressions in $y$ that don't have any common factor and that don't vanish for any $y$ between 0 and 1.

The remaining challenge before we proceed to solve the differential equation \eqref{psiEOM} is to translate the Robin boundary condition for $\Phi$ \eqref{BdyCond} into a boundary condition for $\psi$.  This can be done by writing the first two terms in the series expansion of \eqref{psiDef} at large $\rho$:
 \eqn{psiDefSeries}{
   \Phi (\rho) \sim e^{- i \Omega (\tau + \rho_*)} \left[\psi(1) + {i \Omega \psi(1) \over \rho} - {\rho_H \psi'(1) \over \rho} + {\cal O}\left({1\over \rho^2} \right)\right]\,.
 }
We get:
 \eqn{psiOneBdyCond}{
  \psi'(1) = {1\over \rho_H} \left[{3 \rho_0 \over k_S^2 -2} + i \Omega \right] \psi(1)\,.
 }
Of course, the near horizon boundary condition $V = 0$ translates into
 \eqn{psiZeroBdyCond}{
  \psi(0) = 1\,,
 }
and we can now turn to describing the numerical techniques that we use to solve the differential equation \eqref{psiEOM} with the boundary conditions \eqref{psiOneBdyCond} and \eqref{psiZeroBdyCond}.

\subsection{Numerical method and results}
\label{RESULTS}

Following the method used in \cite{Friess:2006kw} for the computation of quasinormal frequencies of the scalar modes, we integrate the differential equation \eqref{psiEOM} in three steps:  1) we develop a series expansion around $y=0$ and evaluate it at $y=y_i = {1\over 4}$; 2) we integrate the differential equation numerically by using Mathematica's \texttt{NDSolve} from $y = y_i$ to $y = y_f$ (to be given below); and 3) we match our numerical solution onto a series expansion around $y = 1$, which is computed using the boundary condition \eqref{psiOneBdyCond}.  In doing the matching, we compute the Wronskian between the numerical solution and the analytical approximation near $y = 1$.  The Wronskian vanishes only when the two functions are linearly dependent, i.e.\@ when $\psi$ satisfies the boundary condition \eqref{psiOneBdyCond} at $y = 1$.

\begin{figure}
 \centerline{\includegraphics[width=3in]{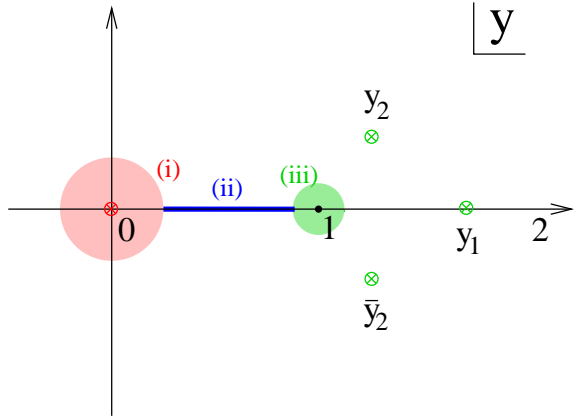}}
  \caption{The zeroes of $s(y)$ represented as crosses in the complex plane.  The red cross at $y=0$ denotes a simple zero, while the green crosses denote double zeroes.  We use a series expansion in region (i), numerical integration in region (ii), and another series expansion in region (iii).} \label{FigPoles}
 \end{figure}

In developing the series expansions, we should keep in mind that the series solutions are guaranteed to converge only when $s(y)$, seen as a function of the complex variable $y$, doesn't vanish.  It is easy to obtain the zeroes of $s(y)$ by writing
 \eqn{sDef}{
  s(y) = y (y-y_1)^2 (y-y_2)^2 (y-\bar{y}_2)^2\,,
 }
where
 \eqn{yiDef}{
  y_1 &= 1 + {k_S^2 - 2 \over 3 (1 + \rho_H^2)}\cr
  y_2 &= 1 + {\rho_H^2 \over 2 (1 + \rho_H^2)} + {i \rho_H \sqrt{4 + 3 \rho_H^2} \over 2 (1 + \rho_H^2)}\,.
 }
It follows that the series expansion around $y = 0$ converges on the whole interval between 0 and 1 (though the convergence close to $y = 1$ might be slow, because of the nearby zero of $s(y)$).  Similarly, the series expansion around $y = 1$ has a radius of convergence $r$ equal to the minimum of $\abs{y_1-1} = {k_S^2 - 2\over 3 (1 + \rho_H^2)}$ and $\abs{y_2-1} = {\rho_H \over \sqrt{1 + \rho_H^2}}$ (see figure~\ref{FigPoles}).  Experience has shown that a good value for $y_f$ was $y_f = 1- r/4$.  

Using the method described above, we computed the lowest nine quasinormal frequencies for $\rho_H=1$ and $l=2$, $3$, $4$, $5$, and $6$ (see table~\ref{RhoHOneOmegas}), and for $\rho_H = 0.2$, $1$, $10$, and $100$ at fixed $l=2$ (see table~\ref{ellTwoOmegas}).  In these tables we have included only the quasinormal modes with $\re \Omega >0$;  equation \eqref{psiEOM} implies that if $\Omega$ is a quasinormal frequency, then so is $-\Omega^*$, so the QNM's with negative real parts can be trivially obtained from the ones with positive $\re \Omega$.

 \begin{table}
  \begin{center}
  \begin{tabular}{|c||c|c|c|c|c|}
   \hline
   $\hbox{freq} \backslash l$ & 2 & 3 & 4 & 5 & 6 \\ \hline\hline
   $\Omega_0$ & $2.156-0.285\,i$ & $3.361-0.354\,i$ & $4.487-0.333\,i$ & $5.561-0.298\,i$ & $6.608-0.266\,i$\\ 
\hline 
$\Omega_1$ & $3.463-2.573\,i$ & $4.461-2.443\,i$ & $5.528-2.271\,i$ & $6.577-2.106\,i$ & $7.610-1.963\,i$\\ 
\hline 
$\Omega_2$ & $5.230-4.942\,i$ & $6.023-4.791\,i$ & $6.964-4.571\,i$ & $7.935-4.340\,i$ & $8.910-4.126\,i$\\ 
\hline 
$\Omega_3$ & $7.096-7.308\,i$ & $7.757-7.165\,i$ & $8.592-6.942\,i$ & $9.484-6.685\,i$ & $10.40-6.432\,i$\\ 
\hline 
$\Omega_4$ & $9.002-9.670\,i$ & $9.572-9.540\,i$ & $10.32-9.327\,i$ & $11.15-9.064\,i$ & $12.00-8.794\,i$\\ 
\hline 
$\Omega_5$ & $10.93-12.03\,i$ & $11.43-11.91\,i$ & $12.12-11.71\,i$ & $12.89-11.45\,i$ & $13.69-11.18\,i$\\ 
\hline 
$\Omega_6$ & $12.86-14.39\,i$ & $13.32-14.28\,i$ & $13.95-14.09\,i$ & $14.68-13.84\,i$ & $15.44-13.56\,i$\\ 
\hline 
$\Omega_7$ & $14.81-16.74\,i$ & $15.23-16.64\,i$ & $15.82-16.47\,i$ & $16.50-16.23\,i$ & $17.23-15.95\,i$\\ 
\hline 
$\Omega_8$ & $16.76-19.10\,i$ & $17.14-19.00\,i$ & $17.70-18.84\,i$ & $18.35-18.61\,i$ & $19.04-18.34\,i$ \\ \hline
  \end{tabular}
  \end{center}
  \caption{Frequencies of scalar quasinormal modes for $\rho_H =1$ in units where $L=1$.}\label{RhoHOneOmegas}
 \end{table}
 \begin{table}
  \begin{center}
  \begin{tabular}{|c||c|c|c|c|}
   \hline
   $\hbox{freq} \backslash \rho_H$ & 0.2 & 1 & 10 & 100 \\ \hline\hline
   $\Omega_0$ & $2.793-0.0008\,i$ & $2.156-0.285\,i$ & $1.739-0.066\,i$ & $1.732-0.007\,i$\\ 
\hline 
$\Omega_1$ & $4.201-0.084\,i$ & $3.463-2.573\,i$ & $18.66-26.63\,i$ & $185.0-266.4\,i$\\ 
\hline 
$\Omega_2$ & $5.468-0.523\,i$ & $5.230-4.942\,i$ & $31.84-49.17\,i$ & $316.1-491.6\,i$\\ 
\hline 
$\Omega_3$ & $6.896-1.121\,i$ & $7.096-7.308\,i$ & $44.95-71.70\,i$ & $446.5-716.8\,i$\\ 
\hline 
$\Omega_4$ & $8.416-1.735\,i$ & $9.002-9.670\,i$ & $58.03-94.22\,i$ & $576.6-941.8\,i$\\ 
\hline 
$\Omega_5$ & $9.984-2.348\,i$ & $10.93-12.03\,i$ & $71.10-116.7\,i$ & $706.6-1167\,i$\\ 
\hline 
$\Omega_6$ & $11.58-2.956\,i$ & $12.86-14.39\,i$ & $84.18-139.2\,i$ & $836.6-1392\,i$\\ 
\hline 
$\Omega_7$ & $13.20-3.559\,i$ & $14.81-16.74\,i$ & $97.25-161.8\,i$ & $966.5-1617\,i$\\ 
\hline 
$\Omega_8$ & $14.83-4.158\,i$ & $16.76-19.10\,i$ & $110.3-184.3\,i$ & $1096-1842\,i$ \\ \hline
  \end{tabular}
  \end{center}
  \caption{Frequencies of scalar quasinormal modes for $l = 2$ in units where $L=1$.}\label{ellTwoOmegas}
 \end{table}
 \begin{table}
  \begin{center}
  \begin{tabular}{|c||c|c|c|c|c|}
   \hline
   $l \backslash \rho_H$ & 5 & 10 & 20 & 50 & 100 \\ \hline\hline
   $2$ & $1.761-0.129\,i$ & $1.739-0.066\,i$ & $1.734-0.033\,i$ & $1.732-0.013\,i$ & $1.732-0.007\,i$\\ 
\hline 
$3$ & $2.547-0.312\,i$ & $2.474-0.164\,i$ & $2.456-0.083\,i$ & $2.450-0.033\,i$ & $2.450-0.017\,i$\\ 
\hline 
$4$ & $3.380-0.537\,i$ & $3.219-0.292\,i$ & $3.177-0.149\,i$ & $3.165-0.060\,i$ & $3.163-0.030\,i$\\ 
\hline 
$5$ & $4.273-0.787\,i$ & $3.979-0.449\,i$ & $3.900-0.231\,i$ & $3.877-0.093\,i$ & $3.874-0.047\,i$\\ 
\hline 
$6$ & $5.230-1.043\,i$ & $4.761-0.631\,i$ & $4.628-0.329\,i$ & $4.590-0.133\,i$ & $4.584-0.067\,i$\\ 
\hline 
$7$ & $6.246-1.286\,i$ & $5.566-0.836\,i$ & $5.362-0.442\,i$ & $5.303-0.180\,i$ & $5.294-0.090\,i$\\ 
\hline 
$8$ & $7.311-1.499\,i$ & $6.399-1.059\,i$ & $6.103-0.570\,i$ & $6.017-0.233\,i$ & $6.004-0.117\,i$\\ 
\hline 
$9$ & $8.410-1.675\,i$ & $7.262-1.297\,i$ & $6.852-0.713\,i$ & $6.731-0.292\,i$ & $6.714-0.147\,i$ \\ \hline
  \end{tabular}
  \end{center}
  \caption{Frequencies of some of the low-lying modes in units where $L=1$.}\label{LowOmegas}
 \end{table}
The most prominent feature of the quasinormal modes included in tables~\ref{RhoHOneOmegas} and \ref{ellTwoOmegas} is the separation of the quasinormal frequencies $\Omega_n$ into two groups:  a main series of fast modes given by $\Omega_n$ with $n\geq 1$, and low-lying slow modes given by $\Omega_0$ (for a similar feature of the quasinormal frequencies in $AdS_5$-Schwarzschild see \cite{Friess:2006kw}).  The low-lying modes differ significantly from the fast ones in a number of ways:
\begin{itemize}
 \item While the fast modes form a tower of modes at each value of $\rho_H$ and $l$, the low-lying modes stand out as not being part of this tower (see figure~\ref{fig:ModesPlot}).
 \begin{figure}
  \centering
  \subfigure[$\rho_H = 0.2$]{\includegraphics[width=3in]{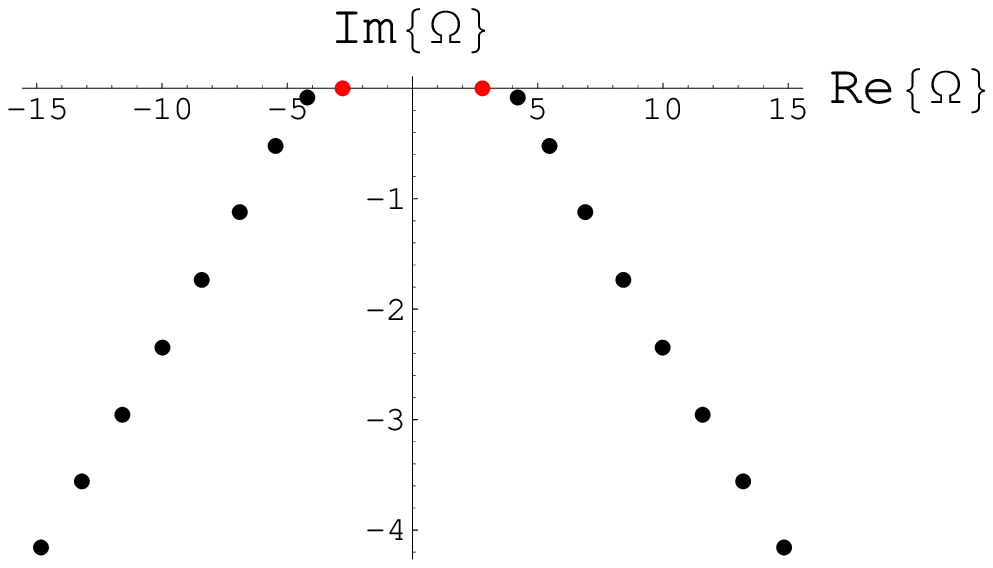}}\qquad
  \subfigure[$\rho_H = 1$]{\includegraphics[width=3in]{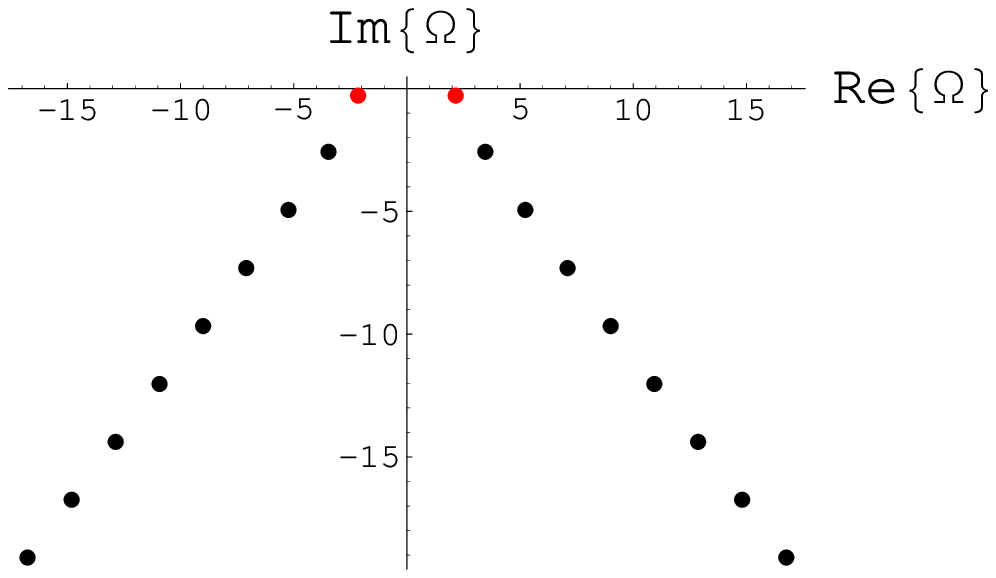}}\\
  \subfigure[$\rho_H = 10$]{\includegraphics[width=3in]{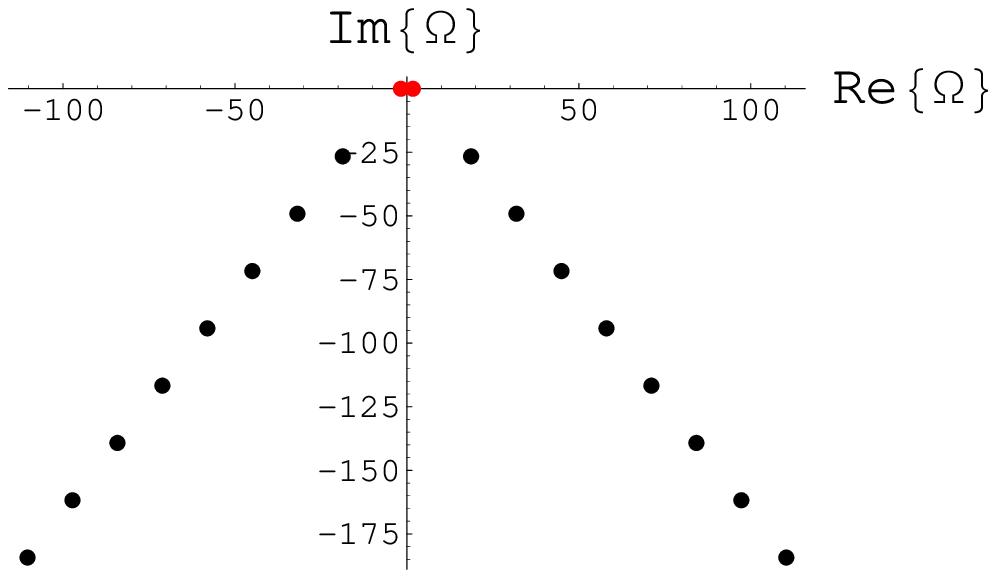}}
  \caption{Quasinormal frequencies for $\rho_H = 0.2$, $\rho_H = 1$, and $\rho_H = 10$, in units where $L=1$.  The black dots represent the main-series modes, while the red ones represent the low-lying modes.  It is fairly clear that for $\rho_H=0.2$ and $1$ the low-lying modes are not part of main series tower.  This is not obvious in the $\rho_H=10$ case, because of the plot scale.}\label{fig:ModesPlot}
 \end{figure}
 \item The low-lying modes have a different $\rho_H$-scaling from the main-series ones (see figure~\ref{fig:ModesPlot}).  This feature is most clearly seen at large $\rho_H$, where the low-lying modes approach $\Omega = \sqrt{l(l+1)}/\sqrt{2}$ as $\rho_H \to \infty$ (see next point), while the main series modes grow proportional to $\rho_H$:  compare, for example, the columns corresponding to $\rho_H = 10$ and $\rho_H = 100$ in table~\ref{ellTwoOmegas}.
 \item The low-lying modes can be interpreted as the linearized hydrodynamic modes of a conformal plasma on $S^2 \times \mathbf{R}$.  While we will explain this correspondence in more detail in section \ref{HYDRO}, for now it is worth mentioning that a linearized hydrodynamics approximation on $S^2 \times \mathbf{R}$ gives, up to first order in $1/\rho_H$, that
  \eqn{HydroPrediction}{
   \Omega = \pm {k_S \over \sqrt{2}} - i {k_S^2 - 2 \over 6 \rho_H} + {\cal O} \left({1\over \rho_H^2} \right)\,,
  }
 with $k_S = \sqrt{l(l+1)}$.  A plot of low-lying modes for various values of $l$ and $\rho_H$, together with the hydrodynamics prediction \eqref{HydroPrediction} can be seen in figure~\ref{fig:LowScalar}, which is based on the numerical values in table~\ref{LowOmegas}.
 \begin{figure}
  \centerline{\includegraphics[width=3in]{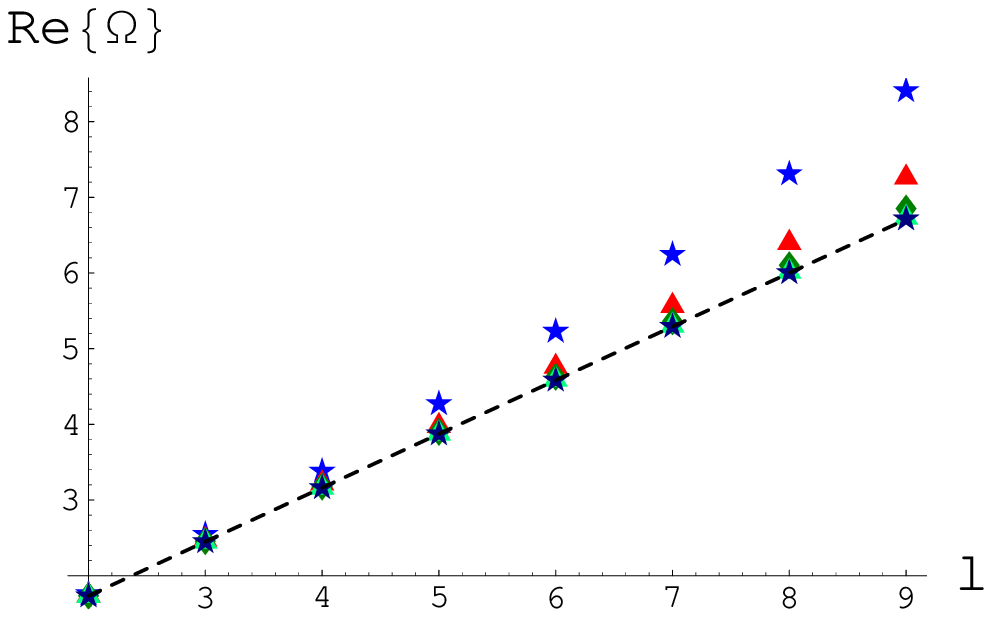}
  \hspace{0.25in}
  \includegraphics[width=3in]{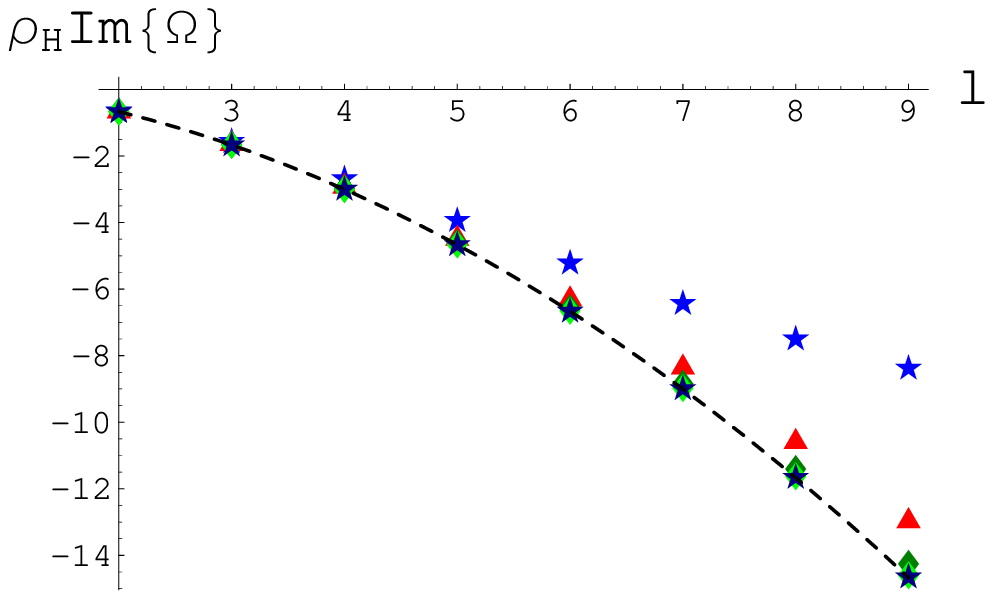}}
  \caption{Quasinormal frequencies for different values of $\rho_H$ plotted against $l$, in units where $L=1$.  The blue stars correspond to $\rho_H = 5$, the red triangles to $\rho_H = 10$, the dark green diamonds to $\rho_H = 20$, the light green triangles (barely visible) to $\rho_H = 50$, and the dark blue stars to $\rho_H = 100$.  The dotted line represents the linearized hydrodynamics prediction \eqref{HydroPrediction}, which matches almost perfectly the numerical results for large $\rho_H$.}\label{fig:LowScalar}
 \end{figure}
\end{itemize}

It is worth noting that while the tower-like feature of the scalar QNM's can be observed even if one imposes a Dirichlet boundary condition, the low-lying modes have \emph{not} been seen in either numerical computations or analytical approximations that use the Dirichlet boundary condition (see, for example, \cite{Cardoso:2003cj, Musiri:2005ev}).

Leaving the low-lying modes aside, we can compare the structure of the main series modes to the structure of the modes described in \cite{Cardoso:2003cj} that come from imposing the Dirichlet boundary condition on the master field.  We find that the spacing between the main series modes at large $\Omega$ asymptotically approaches the spacing between the modes computed in \cite{Cardoso:2001bb}.  However, the initial offset of the tower is different, our modes being in between the ones found in \cite{Cardoso:2001bb}.

\section{Linearized hydrodynamics}\label{HYDRO}

In \cite{Friess:2006kw} it was noticed that in the case of the global $AdS_5$ black hole there was a separation in the imaginary parts of low-lying scalar modes and the ``main series'' modes. The former were interpreted as hydrodynamic modes and the latter as microscopic.  So a simple treatment of linearized hydrodynamics should be able to reproduce these low-lying modes in other dimensions as well.  The goal of this section is to develop such a treatment.

In thinking about hydrodynamics, the general setup on $S^{d-2}\times \mathbf{R}$ is given by the following relations:
 \eqn[c]{Hydro}{
 T_{ab} = (\epsilon+p)u_{a}u_{b} + p \tilde{g}_{ab} + \tau_{ab} \cr
 \tau_{ab}  \equiv  - \eta \left( \Delta_{ac}\tilde{\nabla}^{c}u_{b} + \Delta_{bc}\tilde{\nabla}^{c}u_{a} -\frac{2}{d-2}\Delta_{ab}\tilde{\nabla}^{c}u_{c} \right) - \xi \Delta_{ab}\tilde{\nabla}^{c}u_{c} \cr
 \Delta_{ab} = \tilde{g}_{ab} + u_{a}u_{b} \cr
 \tilde{\nabla}^{a}T_{ab} =0
 \; .}
where $\tilde{g}_{ab}$ is the metric on $S^{d-2}\times \mathbf{R}$ and $\tilde{\nabla}_a$ is the covariant derivative with respect to this metric.  Since the theory on the boundary of AdS is conformal we expect  $T^{a}_{\phantom{a}a} =0$,  which implies both $\epsilon = (d-2) p $  and $ \xi =0$.   Following the same approach as in \cite{Friess:2006kw}, we ignore the temperature-independent contribution from the Casimir energy to $T^{ab}$.  The Casimir energy comes from considering the quantum field theory on the compact space $S^{d-2}$.  For our purposes we can think of it as a temperature-independent shift of the zero point energy, which can be safely ignored. 

The vector $u^{a}$ describes the velocity at each point in the fluid, and we choose to normalize it by imposing $u^{a}u_{a} =-1$.  Let us denote $u^{a} = (1, u^{i})$  where $i$ runs over the $ S^{d-2}$ directions. In the linearized approximation we consider $u^{i}$ to be small.  Perturbing at the same time the  pressure $ p =p_{0} + \delta p $, one can derive from \eno{Hydro} the linearized equations
\eqn[c]{LinearHydro}{ (d-2) \frac{\partial \delta p}{\partial \tau} + (d-1) p_{0} \tilde{\nabla} _{i} u^{i}  =0 \cr
(d-1) p_{0} \frac{\partial u^{i}}{\partial \tau} + \tilde{\nabla}^{i} \delta p - \eta\left( \tilde{\nabla}^{j} \tilde{\nabla}_{j} u^{i} +\tilde{\nabla}_{j} \tilde{\nabla}^{i} u^{j}\right) - \eta \frac{2}{d-2}\partial^{i} \tilde{\nabla}_{j}u^{j} =0
 \; . }
Note that for $d = 5$ equation \eqref{LinearHydro} reduces to the linearized Navier-Stokes equation on $S^{3}$, which was analyzed in section 5.3 of \cite{Friess:2006kw}.  We wish to examine scalar perturbations next, which are described by the ansatz
\eqn[c]{HydScalAnsatz}{\delta p = K_{1} e^{-i\Omega \tau} \mathbb{S} \qquad u^{i} = K_{2} e^{-i \Omega \tau} \tilde{\nabla}^{i} \mathbb{S}\,,
}
where $\mathbb{S}$ satisfies $ \left( \tilde{\nabla}^{2} +k_{S}^2\right) \mathbb{S} = 0$, as explained in section~\ref{SETUP}, and $L=1$.  Plugging \eqref{HydScalAnsatz} into \eqref{LinearHydro}, we obtain the following system of equations for $K_{1}$ and $K_{2}$:
\eqn[c]{KDet}{-i\Omega (d-2) K_{1} - (d-1)k_{S}p_{0}K_{2}  =0 \cr
-k_{S}K_{1} + \left(-i \Omega(d-1)p_{0} +2\eta \frac{d-3}{d-2}k_{S}^{2}-2(d-3)\eta\right)K_{2} =0\,.
}
In order to have non-trivial solutions, this system must have zero determinant.  This gives a quadratic equation for $\Omega$, whose solutions can be given in terms of a series expansion in $\eta/p_0$:
\eqn{GotHydScalOmega}{ \Omega = \pm \frac{k_{S}}{\sqrt{d-2}} - i \frac{\eta}{p_{0}}\frac{k_{S}^{2}(d-3) -(d-2)(d-3)}{(d-1)(d-2)} +\mathcal{O}\left(\frac{\eta^2}{p_{0}^{2}}\right)
}
We can connect this result to the $AdS_{4}$ quasinormal mode problem by noting that
\eqn{etaoverp0}{
  {\eta \over p_0} = {4 \pi \eta \over s} {\rho_H \over 1 + \rho_H^2}\,,
 }
which can be easily derived from \eno{MET} in the case $d=4$, but it is actually true in any number of dimensions.  Using the conjectured lower bound on the viscosity $\frac{\eta}{s} =\frac{1}{4\pi}$ \cite{Kovtun:2004de,Policastro:2001yc}, that has been checked in the $AdS_{4}$ case in \cite{Herzog:2002fn}, we find
\eqn{HydScalOmegaRhoH}{\Omega = \pm \frac{k_{S}}{\sqrt{d-2}}- i \frac{1}{\rho_{H}} \frac{k_{S}^{2}(d-3) -(d-2)(d-3)}{(d-1)(d-2)} + \mathcal{O} \left(\frac{1}{\rho_{H}^{2}}\right)\,.}
It is easily seen that this reproduces the hydrodynamical modes previously discussed in the global $AdS_{5}$ black hole case in \cite{Friess:2006kw}.  For $d=4$, equation \eqref{HydScalOmegaRhoH} reduces to \eno{HydroPrediction}.

Similarly, we can describe the low-lying vector modes by the ansatz:
\eqn{Hyd}{\delta p =0 \qquad u^{i} = K_{3} e^{-i \Omega \tau} V^{i}
\; . }
We find that the corresponding frequencies $\Omega$ are given by
\eqn[c]{GotHydVecOmega}{ \Omega = -i (d-1) \frac{ 1 - \sqrt{1- 4 k_{V}^{2} \frac{\eta^{2}}{(d-1)^2 p_{0}^{2}}}}{2\eta /p_{0}} = -i \frac{k_{V}^{2} \eta}{(d-1) p_{0}} + \mathcal{O}\left(\frac{\eta ^{2}}{p_{0}^{2}}\right) \cr \Omega = -i \frac{k_{V}^{2}}{d-1} \frac{1}{\rho_{H}} + \mathcal{O}\left( \frac{1}{\rho_{H}^{2}}\right) \;. }
It is interesting to note that the numerical value given by this formula when $d=4$, $l =2$ and $\rho_{H} =100$ agrees within $10 \%$ with the low-lying vector mode of Table $9$ in \cite{Cardoso:2003cj}.

\section{Conclusions}\label{CONCLUSIONS}

In this note we examine the relation between the asymptotic behavior of the master field and the behavior of the scalar sector of metric perturbations in the global $AdS_{4}$ black hole.  We argue that the boundary condition that corresponds to a non-deformation of the metric on the boundary translates into a Robin boundary condition for the master field.  With this boundary condition, we compute the scalar quasinormal modes.  We find some low-lying modes that have not been previously observed, and compare them with the linearized hydrodynamical modes of the boundary CFT. 
\section*{Acknowledgements}\label{ACKNOWLEDGEMENTS}
We would like to thank S.~Gubser for suggesting this problem, for valuable discussions, and a careful reading of the draft.

\clearpage
\appendix
\section{Gauge freedom and boundary conditions}
\label{APPENDIX}
In this section we derive the asymptotic expressions for the metric perturbations $\delta g_{ab}$ by solving the system of equations \eqref{ScalarPerturb}--\eqref{ScalarEOM}.  In doing so, it is important to realize that equations \eqref{ScalarPerturb}--\eqref{ScalarEOM} don't determine the metric perturbations uniquely.  We have seen that the gauge freedom\footnote{In this section $\nabla_a$ denotes the covariant derivative with respect to the four-dimensional metric \eno{Metric}.}
 \eqn{GaugeFreedom}{
  \delta g_{ab} \to \delta g_{ab} + \nabla_a v_b + \nabla_b v_a
 }
present in any perturbation theory problem in general relativity enables us to set $\delta g_{\rho a} = 0$ (this is what we referred to as axial gauge).  However, even after we set $\delta g_{a \rho} = 0$ we still have a residual gauge freedom left, and we would like to understand this residual gauge freedom a bit better before we derive the asymptotic expressions for $\delta g_{ab}$.

The first thing to note is that generic gauge transformations of type \eqref{GaugeFreedom} do not preserve the form of the metric \eqref{ScalarPerturb}.  Instead, generic transformations would just map our initial solution onto something that doesn't transform under the $SO(3)$ isometry group of $S^2$ in any definite way.  We only look at  perturbations with the specific $SO(3)$ structure defined in \eqref{ScalarPerturb}.  Therefore, we need to restrict the class of allowed gauge transformations to the ones that preserve this $SO(3)$ structure.  Such transformations are of the form
 \eqn{AllowedTransf}{
  v_a(\tau, \rho, \theta, \phi) = \begin{pmatrix} h_\tau(\tau, \rho) \mathbb{S}(\theta, \phi) & h_\rho(\tau, \rho) \mathbb{S}(\theta, \phi) & h(\tau, \rho) \mathbb{S}_\theta(\theta, \phi) &  h(\tau, \rho) \mathbb{S}_\phi(\theta, \phi)\end{pmatrix}\,,
 }
and they give
 \eqn[c]{TransfStructure}{
  2 \nabla_{(\alpha} v_{\beta)} = 2 D_{(\alpha} h_{\beta)} \mathbb{S} \qquad 2 \nabla_{(\alpha} v_{i)} = \left[\partial_\alpha h - h_\alpha k_S - {2 \over \rho} (\partial_\alpha \rho) h \right] \mathbb{S}_i\cr
  2 \nabla_{(i} v_{j)} = -2 h k_S \mathbb{S}_{ij} + \left(h k_S + 2 \rho f h_\rho \right) \gamma_{ij} \mathbb{S} \,.
 }
It is easy to see now that if we start with any scalar perturbation of the form \eqref{ScalarPerturb}, we can set $\delta g_{\rho a} = 0$ by solving three first order non-homogeneous differential equations for $h_\tau$, $h_\rho$, and $h$.  The residual freedom that remains after setting $\delta g_{\rho a} = 0$ is reflected in the choice of the three integration constants (which are functions of $\tau$) that enter in the general solutions of these equations.  So in addition to setting $\delta g_{\rho a} = 0$ we also have the freedom to prescribe the time behavior of \emph{three} of the other components of $\delta g_{ab}$ at a given point.  In particular, the gauge freedom allows us to set the large $\rho$ behavior of three such components to have no $\rho^2$ term in a large $\rho$ series expansion.  Requiring all of these components (which are described by the \emph{four} functions $H_{T}$, $H_{L}$, $f_{\tau}$, $f_{\tau \tau}$) to have no $\rho^2$ terms cannot be accomplished by making a gauge transformation, and is therefore meaningful as a boundary condition on the metric perturbations.

We now turn to the problem of finding the asymptotic expressions for the metric coefficients $\delta g_{ab}$ in axial gauge.  We will set $L=1$ throughout the entire calculation.  With the assumption
 \eqn{Assumptions}{
  f_{\rho \rho} &= f_{\rho} = 0 \qquad f_{\tau\tau} = e^{-i \Omega \tau} B(\rho) \qquad f_{\tau} = e^{-i \Omega \tau} C(\rho)\cr
  H_L &= e^{-i \Omega \tau} A_L(\rho) \qquad \qquad H_T = e^{-i \Omega \tau} A_T(\rho)
 }
we first compute the quantities $F$ and $F_{\alpha\beta}$ that enter in equations \eqref{ScalarPhiProperties} and \eqref{ScalarPhi}:
 \eqn{FEquations}{
  F &= {e^{-i \Omega \tau} \over 2 k_S^2} \left(k_S^2 \left[2 A_L(\rho) + A_T(\rho) \right] + 2 \rho f A_T'(\rho) \right)\cr
  F_{\tau\tau} &= {e^{-i \Omega \tau} \over k_S^2} \left(-2 i k_S \rho \Omega C(\rho) +\\ k_S^2 B(\rho) - \rho^2 \left[2 \Omega^2 A_T(\rho) + f f' A_T'(\rho) \right] \right)\cr
  F_{\tau\rho} &= {e^{-i \Omega \tau} \over k_S^2 f} \big( k_S f C(\rho) -2 i \Omega \rho f \left[A_T(\rho) + \rho A_T'(\rho) \right] +\cr &\qquad\qquad\qquad\qquad\qquad + \rho f k_S C'(\rho) + i \Omega \rho^2 f' A_T(\rho) - k_S \rho f' C(\rho) \big)\cr
  F_{\rho\rho} &= {e^{-i \Omega \tau} \rho \over k_S^2 f} \left(A_T'(\rho) \left[4 f + \rho f' \right] + 2 \rho f A_T''(\rho) \right)\,,
 }
and $F_{\rho\tau} = F_{\tau\rho}$.  The plan now is to plug the above expressions into equations \eqref{ScalarPhiProperties} and \eqref{ScalarPhi}, and find a series solution for the corresponding differential equations.  In order to do this though, we need to get hold of the right-hand side of equation \eqref{ScalarPhi}, perhaps in the form of a large $\rho$ series expansion.  This can be done by solving the master equation \eqref{ScalarEOM}:
 \eqn{PhiSeries}{
  \Phi(\rho) = \varphi^{(0)} + {\varphi^{(1)} \over \rho} + {\varphi^{(2)} \over \rho^2} + {\varphi^{(3)} \over \rho^3} + \dotsb\,,
 }
where
 \eqn{PhiSeriesDefs}{
  \varphi^{(2)} &= \left[ {9 \rho_0^2 \over (k_S^2 -2)^2} + {k_S^2 - \Omega^2 \over 2} \right] \varphi^{(0)}\cr
  \varphi^{(3)} &= \left[-{18 \rho_0^3 \over (k_S^2-2)^3} - {\rho_0 (2 + k_S^2)\over 2(k_S^2 -2)} \right] \varphi^{(0)} + \left[{3 \rho_0^2 \over (k_S-2)^3} + { k_S^2 -2 -\Omega^2 \over 6}  \right] \varphi^{(1)}
 }
and all higher order terms can be expressed in terms of linear combinations of $\varphi^{(0)}$ and $\varphi^{(1)}$.  The two constants $\varphi^{(0)}$ and $\varphi^{(1)}$ can thus be interpreted as the two integration constants that appear when we integrate the master equation, which is a second order ODE.  The above expansion can then be used to find a series expansion of the right-hand side of equation \eqref{ScalarPhi}.  The resulting expressions are long and not that insightful, so we will not reproduce them here;  their derivation is nevertheless straightforward.

We now solve for the functions $A_L(\rho)$, $A_T(\rho)$, $B(\rho)$, and $C(\rho)$ that completely determine the metric perturbations via
 \eqn[c]{ScalarPerturbAgain}{
  \delta g_{\rho\rho} = \delta g_{\tau\rho} =
   \delta g_{\rho i} = 0 \cr
  \delta g_{\tau\tau} = e^{-i \Omega \tau} B(\rho) \,
    \mathbb{S}(\theta,\phi) \qquad
   \delta g_{\tau i} = \rho e^{-i \Omega \tau} C(\rho) \,
     \mathbb{S}_i(\theta,\phi)  \cr
   \delta g_{ij} = 2 \rho^2 e^{-i \Omega \tau} \left[ A_L(\rho) \, \gamma_{ij} \,
    \mathbb{S}(\theta,\phi) + A_T(\rho) \,
    \mathbb{S}_{ij}(\theta,\phi) \right]
 }
in four steps:
\begin{enumerate}
  \item We first solve for $A_T(\rho)$ from the $F_{\rho\rho}$ equation in \eqref{ScalarPhi} with the LHS given by the corresponding expression in \eqref{FEquations} and the RHS computed from the series expansion \eqref{PhiSeries}.  We find:
   \eqn{ATDef}{
    A_T(\rho) = A_T^{(0)} + {A_T^{(2)} \over \rho^2} + \Bigg[ {k_S^2 \rho_0 \over 2(k_S^2-2)}& \left(\varphi^{(1)} + {3 \rho_0 \varphi^{(0)} \over k_S^2-2} \right) +\cr
    &+ {k_S^2 (k_S^2 - 2 \Omega^2) \over 12} \varphi^{(0)} \Bigg] {1\over \rho^3} + {\cal O} \left({1\over \rho^4} \right)\,.
   }
  Here, we can think of $A_T^{(0)}$ and $A_T^{(2)}$ as integration constants:  we have two integration constants because the differential equation satisfied by $A_T(\rho)$ is second order, as can be easily seen from the $F_{\rho\rho}$ relation in \eqref{FEquations}.
  \item Next, we solve for $C(\rho)$ from the $F_{\tau\rho}$ equation in \eqref{ScalarPhi}.  Again, the LHS of this equation is given in \eqref{FEquations}, and the RHS can be computed from \eqref{PhiSeries}.  We obtain:
   \eqn{CDef}{
    C(\rho) = C^{(-1)} \rho &+ \left[-{i k_S \Omega\over 2} \left(\varphi^{(1)} + {3 \rho_0 \varphi^{(0)} \over k_S^2-2} \right)  + {k_S C^{(-1)} +i \Omega (2 A_T^{(2)} - A_T^{(0)}) \over k_S}\right] {1\over \rho}+\cr
    &+ \left[{\rho_0 (i \Omega A_T^{(0)}- k_S C^{(-1)}) \over k_S} -{i k_S \Omega (k_S^2-2) \over 6} \varphi^{(0)} \right] {1\over \rho^2} + {\cal O}\left({1\over \rho^3} \right)\,,
   }
  where $C^{(-1)}$ is an integration constant---we can see from the $F_{\tau\rho}$ relation in \eqref{FEquations} that the corresponding differential equation for $C(\rho)$ is a first order ODE, so its solution has to have one integration constant.
  \item Similarly, we next solve for $B(\rho)$ from the $F_{\tau\tau}$ equation in \eqref{ScalarPhi}.  As can be seen from the $F_{\tau\tau}$ relation in \eqref{FEquations}, this equation doesn't involve any derivatives of $B(\rho)$, so its solution doesn't involve additional integration constants:
   \eqn{BDef}{
    B(\rho) &= {2 (i k_S \Omega C^{(-1)} + \Omega^2 A_T^{(0)} -2 A_T^{(2)}) \over k_S^2} \rho^2 + \Bigg[{k_S^2 - 2 + 2 \Omega^2 \over 4 (k_S^2 - 2)} \left(\varphi^{(1)} + {3 \rho_0 \varphi^{(0)} \over k_S^2-2} \right) +\cr
    &+ {2 \Omega (i k_S C^{(-1)} + \Omega A_T^{(0)} ) - 2 (1 + \Omega^2) A_T^{(2)}\over k_S^2} \Bigg] + \Bigg[\rho_0\left(\varphi^{(1)} + {3 \rho_0 \varphi^{(0)} \over k_S^2-2} \right) -\cr
    &-{2 \rho_0 \Omega (i k_S C^{(-1)} + \Omega A_T^{(0)}) \over k_S^2} + {k_S^2 (k_S^2-2)\over 6} \varphi^{(0)} \Bigg]{1\over \rho} + {\cal O}\left({1\over \rho^2}\right)\,.
   }
  \item Finally, we solve for $A_L(\rho)$ from the second equation in \eqref{ScalarPhiProperties} with $\beta = \tau$.  Again, this equation doesn't involve any derivatives, so we have no integration constants:
   \eqn{ALDef}{
    A_L(\rho) &= \Bigg[{4 A_T^{(2)} - k_S^2 A_T^{(0)} \over 2 k_S^2} - {1\over 2}\left(\varphi^{(1)} + {3 \rho_0 \varphi^{(0)} \over k_S^2-2} \right)\Bigg] +\cr
    &+ \Bigg[-{(k_S^2 - 2) A_T^{(2)} \over 2 k_S^2} + {k_S^2 -2 \over 8}\left(\varphi^{(1)} + {3 \rho_0 \varphi^{(0)} \over k_S^2-2} \right) \Bigg] {1\over \rho^2} +\cr
    &+ \Bigg[-{\rho_0 A_T^{(2)} \over k_S^2} + {\rho_0 \over 4} \left(\varphi^{(1)} + {3 \rho_0 \varphi^{(0)} \over k_S^2-2} \right) + {k_S^2 (k_S^2-2) \over 24} \varphi^{(0)} \Bigg] {1\over \rho^3} + {\cal O}\left({1\over \rho^4} \right)\,.
   }
\end{enumerate}

It can be checked that the above series solutions automatically satisfy the other two equations in \eqref{ScalarPhiProperties} that we have not used.  Also, the fact that the integration constants $A_T^{(0)}$, $A_T^{(2)}$, and $C^{(-1)}$ are still undetermined is a consequence of the residual gauge freedom that we discussed at the beginning of this section:  these three integration constants allow us to set the values of \emph{three} of the functions $A_L(\rho)$, $A_T(\rho)$, $B(\rho)$, and $C(\rho)$ at a given point to whatever we want.

 Requiring that the metric perturbations don't grow like $\rho^2$ at large $\rho$ and  using  \eqref{ScalarPerturbAgain}--\eqref{ALDef}, we get  $A_T^{(0)}=A_T^{(2)}=C^{(-1)}=0$, together with
 \eqn{BdyCondAgain}{
  \varphi^{(1)} + {3 \rho_0 \varphi^{(0)} \over k_S^2-2} = 0\,,
 }
which is the same as \eqref{BdyCond}.

We note that the relations \eqref{BdyCondAgain} and $A_T^{(0)}=A_T^{(2)}=C^{(-1)}=0$ make almost all terms written in the series expansions \eqref{ATDef}--\eqref{ALDef} disappear, and we're left just with
 \eqn{WhatsLeft}{
  H_L &= {k_S^2 (k_S^2-2) \over 24} e^{-i \Omega \tau} {\varphi^{(0)}  \over \rho^3} + {\cal O}\left({1\over \rho^4} \right)\cr
  H_T &= {k_S^2 (k_S^2 - 2 \Omega^2) \over 12} e^{-i \Omega \tau} {\varphi^{(0)} \over \rho^3} + {\cal O} \left({1\over \rho^4} \right)\cr
  f_{\tau\tau} &= {k_S^2 (k_S^2-2) \over 6} e^{-i \Omega \tau} {\varphi^{(0)} \over \rho} + {\cal O}\left({1\over \rho^2}\right)\cr
  f_{\tau} &= -{i k_S \Omega (k_S^2-2) \over 6} e^{-i \Omega \tau} {\varphi^{(0)} \over \rho^2} + {\cal O}\left({1\over \rho^3} \right)\,,
 }
which looks incredibly similar to the expressions found in section 3.3.2 of \cite{Friess:2006kw} in $AdS_5$-Schwarzschild.  In light of the analysis done in \cite{Friess:2006kw}, it is worth mentioning that the leading coefficients in \eqref{WhatsLeft} give, up to a proportionality factor, the expectation value of the stress-energy tensor in the boundary $2+1$-dimensional CFT.  Conservation and tracelessness of the stress-energy tensor can then be easily checked using the same approach as in \cite{Friess:2006kw}.

\clearpage
\bibliographystyle{ssg}
\bibliography{lowmodes}

\end{document}